\documentclass[runningheads]{llncs}
\usepackage[T1]{fontenc}
\usepackage{graphicx}
\usepackage{color}
\usepackage{xurl}
\usepackage{hyperref}
\usepackage{wrapfig}
\usepackage{acronym, soul}
\usepackage{multirow}
\usepackage{enumitem}
\usepackage{subfigure}
\usepackage{bm}
\usepackage{bbding}
\usepackage{diagbox}
\usepackage{slashbox}
\usepackage{booktabs, soul}
\usepackage{amsmath}
\usepackage{array, makecell}
\definecolor{redtable}{RGB}{255, 153, 153}
\definecolor{greentable}{RGB}{128, 255, 128}
\newacro{ai}[AI]{artificial intelligence}
\newacro{ae}[AE]{autoencoder}
\newacro{auroc}[AUROC]{area under the receiver operating characteristic}
\newacro{auprc}[AUPRC]{area under the precision-recall curve }
\newacro{btae}[B\_TAE]{basic transformer autoencoder}
\newacro{brats}[BraTS]{brain tumor segmentation}
\newacro{crf}[CRF]{conditional random field}
\newacro{ct}[CT]{computed tomography}
\newacro{cnn}[CNN]{convolutional neural network}
\newacro{cpu}[CPU]{central processing unit}
\newacro{dsc}[DSC]{dice similarity coefficient}
\newacro{dl}[DL]{deep learning}
\newacro{dctae}[DC\_TAE]{dense convolutional transformer autoencoder}
\newacro{elbo}[ELBO]{evidence lower bound objective}
\newacro{uad}[UAD]{unsupervised anomaly detection}
\newacro{unet}[U-Net]{U-shaped network}
\newacro{flair}[FLAIR]{fluid attenuated inversion recovery}
\newacro{fcnn}[FCNN]{fully convolutional neural network}
\newacro{gan}[GAN]{generative adversarial network}
\newacro{gpu}[GPU]{graphical processing unit}
\newacro{gru}[GRU]{gated recurrent units}
\newacro{gb}[GB]{Glioblastoma}
\newacro{htae}[H\_TAE]{hierarchical transformer autoencoder}
\newacro{htaes}[H\_TAE\_S]{hierarchical transformer autoencoder skip connections}
\newacro{irm}[IRM]{{\it imagerie par résonance magnétique}}
\newacro{lvm}[LVM]{latent variable model}
\newacro{lstm}[LSTM]{long short-term memory}
\newacro{mae}[MAE]{mean absolute error}
\newacro{mse}[MSE]{mean squared error}
\newacro{mri}[MRI]{magnetic resonance imaging}
\newacro{mr}[MR]{magnetic resonance}
\newacro{ml}[ML]{machine learning}
\newacro{ms}[MS]{multiple sclerosis}
\newacro{mc}[MC]{Monte Carlo}
\newacro{mslub}[MSLUB]{multiple sclerosis lesion}
\newacro{mood}[MOOD]{medical out-of-distribution}
\newacro{miccai}[MICCAI]{medical image computing and computer-assisted intervention}
\newacro{mrf}[MRF]{Markov random field}
\newacro{nlp}[NLP]{natural language processing}
\newacro{nifti}[NIfTI]{neuroimaging informatics technology initiative}
\newacro{ocsvm}[OC-SVM]{one-class support vector machine}
\newacro{oct}[OCT]{optical coherence tomography}
\newacro{oasis}[OASIS]{open access series of imaging studies}
\newacro{pet}[PET]{positron emission tomography}
\newacro{pca}[PCA]{principal component analysis}
\newacro{prnn}[PixelRNN]{pixel recurrent neural network}
\newacro{resnet}[ResNet]{residual neural network}
\newacro{rnn}[RNN]{recurrent neural network}
\newacro{roc}[ROC]{receiver operating characteristics}
\newacro{sctae}[SC\_TAE]{spatial convolutional transformer autoencoder}
\newacro{sota}[SOTA]{state-of-the-art}
\newacro{ssim}[SSIM]{structural similarity image metric}
\newacro{tae}[TAE]{transformer autoencoder}
\newacro{tnn}[TNN]{transformer neural network}
\newacro{vit}[ViT]{vision transformer}
\newacro{vae}[VAE]{variational autoencoder}
\newacro{vqvae}[VQ-VAE]{vector quantizer variational autoencoder}
\newacro{vaegan}[VAEGAN]{variational autoencoder generative adversarial network}
\newacro{vpn}[VPN]{virtual private network}
\newacro{knn}[KNN]{k-nearest neighbors}
\newacro{kg}[KG]{knowledge-guided}
\newacro{ks}[KS]{Kolmogorov–Smirnov}
\begin{document}
\title{Transformer based Models for Unsupervised Anomaly Segmentation in Brain MR Images}
\author{Ahmed Ghorbel\inst{1}\orcidID{0000-0003-4213-769X}
\and
Ahmed Aldahdooh\inst{1}\orcidID{0000-0002-7624-5253}
\and
Shadi Albarqouni\inst{2}\orcidID{0000-0003-2157-2211}
\and
Wassim Hamidouche\inst{1}\orcidID{0000-0002-0143-1756}}
%
\institute{Univ. Rennes, INSA Rennes, CNRS, IETR - UMR 6164, 20 Av. des Buttes de Coesmes, 35700, Rennes, France \\
\email{\{firstName.lastName\}@insa-rennes.fr} \\ 
\url{https://www.ietr.fr/} 
\url{https://www.insa-rennes.fr/} 
\and
University Hospital Bonn, Venusberg-Campus 1, D-53127, Bonn, Germany\\
Helmholtz Munich, Ingolstädter Landstraße 1, D-85764, Neuherberg, Germany\\
Technical University of Munich, Boltzmannstr. 3, D-85748 Garching,  Germany\\
\email{shadi.albarqouni@ukbonn.de}}
\maketitle

\begin{abstract}
The quality of patient care associated with diagnostic radiology is proportionate to a physician’s workload. Segmentation is a fundamental limiting precursor to both diagnostic and therapeutic procedures. Advances in \ac{ml} aim to increase diagnostic efficiency by replacing a single application with generalized algorithms. The goal of \ac{uad} is to identify potential anomalous regions unseen during training, where \ac{cnn} based \acp{ae}, and \acp{vae} are considered a de facto approach for reconstruction based-anomaly segmentation. The restricted receptive field in \acp{cnn} limits the \ac{cnn} to model the global context. Hence, if the anomalous regions cover large parts of the image, the \ac{cnn}-based \acp{ae} are not capable of bringing a semantic understanding of the image. Meanwhile, \acp{vit} have emerged as a competitive alternative to \acp{cnn}. It relies on the self-attention mechanism that can relate image patches to each other. We investigate in this paper Transformer's capabilities in building \acp{ae} for the reconstruction-based \ac{uad} task to reconstruct a coherent and more realistic image. We focus on anomaly segmentation for brain \ac{mri} and present five Transformer-based models while enabling segmentation performance comparable to or superior to \ac{sota} models. The source code is made publicly available on \href{https://github.com/ahmedgh970/Transformers_Unsupervised_Anomaly_Segmentation.git}{GitHub}.
\end{abstract}
%
%
%
\section{Introduction}
In recent years, significant progress has been achieved in developing deep learning approaches for tackling various tasks. Anomaly segmentation is one of the most challenging tasks in computer vision applications. It seeks to act like experienced physicians to identify and delineate various anomalies on medical images, such as tumors on brain \ac{mr} scans. Therefore, it is crucial to constantly improve the accuracy of medical image segmentation by developing novel \ac{dl} techniques. Since the advent of \ac{dl}, \ac{fcnn} and in particular U-shaped autoencoder architectures \cite{isensee2021nnu,siddique2021u} have achieved \ac{sota} results in various medical semantic segmentation tasks. Additionally, in computer vision, using Transformers as a backbone encoder is beneficial due to their great capability of modeling long-range dependencies and capturing global context \cite{https://doi.org/10.48550/arxiv.2010.11929}.

Detecting and diagnosing diseases, monitoring illness development, and treatment planning are all common uses for neuroimaging.
Manually identifying and segmenting diseases in the brain \ac{mri} is time-consuming and arduous.
The medical image analysis community has offered a wide range of strategies to aid in detecting and delineating brain lesions emerging from \ac{ms} lesions or tumors.
Many supervised \ac{dl} techniques have attained outstanding levels of performance like supervised \ac{knn} and automatic \ac{kg} for glioma patients and supervised CNN-based methods for multiple sclerosis lesion segmentation described throughout this survey \cite{zhang2020multiple}.
Nevertheless, these techniques, which are primarily based on supervised \ac{dl}, have certain drawbacks:
1) A relevant pathology is estimated to be missed in 5-10\% of scans \cite{bruno2015understanding}. 
2) Training supervised models requires large and diverse annotated datasets, which are scarce and expensive to acquire.
3) The resulting models are limited to detecting lesions similar to those seen in the training data.

Furthermore, these methods have a limited generalization since training data rarely covers the whole range of clinical manifestations  \cite{taboada2009anomaly}. To overcome these challenges, recent works~\cite{baur2021autoencoders,baur2021modeling} investigated modeling the distribution of healthy brains to detect pathologies as a deviation from the norm. The challenge of brain anomalies identification and delineation is formulated here as an \ac{uad} task based on \ac{sota} deep representation learning, which requires only a set of healthy data. Anomalies are detected and delimited first by computing the pixel-wise $L_1$-distance between an input image and its reconstruction, and then post-processing the resulting residual image to get a binary segmentation, as illustrated in Fig.~\ref{fig_process} and Fig.~\ref{fig1} of Appendix~\ref{appA}.

Due to the inherent locality of convolution processes, CNN-based techniques have difficulties modeling explicit long-range relationships, despite their outstanding representational capability. Therefore, these architectures generally yield weak performance, especially for target structures that show significant inter-patient variation in texture, shape, and size. 
On the other hand, Transformers exempt convolution operators entirely and rely solely on attention mechanisms to capture the interactions between inputs, regardless of their relative position to one another, like \acp{prnn} \cite{van2016pixel}. Unlike prior CNN-based methods known for creating blurry reconstructions, Transformers are not only effective in modeling global contexts but also show greater transferability for downstream tasks under large-scale pre-training.
In this work, we propose a method for unsupervised anomaly segmentation using Transformers, where we learn the distribution of healthy brain \ac{mr} data. We create and evaluate different Transformer-based models and compare their performance results on real anomalous datasets with recent \ac{sota} unsupervised models \cite{baur2021autoencoders}. 

Although Transformers need a large-scale dataset for training, we train our models and SOTA models from scratch on the healthy OASIS-3~\cite{lamontagne2019oasis} dataset, and we don't use extra data for pre-trained models. Two challenging datasets MSLUB~\cite{lesjak2018novel} and BraTS-20~\cite{menze2014multimodal} are tested for detecting two different pathologies.  To summarize, our contributions are mainly focusing on building Transformer-based models to answer the following questions:
\begin{itemize}[noitemsep,topsep=4pt,itemsep=2pt,partopsep=4pt, parsep=4pt, leftmargin=12pt]
\item Is the basic Transformer architecture capable of modeling the distribution of healthy data? Due to the Transformer's ability to model the global context of the image, we found that it can reconstruct anomalous inputs with the same intensity levels. Thus it fails in modeling the healthy data distribution for the anomaly segmentation task. 
\item Are the joint Transformer and CNN architectures capable of modeling the distribution of healthy data? Due to the ability to model global and local features in the common architecture, we found that having Dense Convolutional \ac{ae} between the encoder and the decoder of the Transformer yields superior performance compared to CNN-based \acp{ae} in BraTS dataset and comparable performance in MSLUB dataset. 
\item Is the hierarchical Transformer architecture able to model the latent space of the healthy data? Owing to the ability to capture short/long-range spatial correlations by down/up-sampling and combining the self-attention provided by  Transformer layers, respectively,  we can say yes, and it yields superior performance in BraTS dataset compared to other \ac{cnn} and Transformer based architectures. 
\end{itemize}

\section{Related Works}
\textbf{CNN-based \ac{uad} Networks:} Since the introduction of \acp{ae} \cite{baur2018deep} and their generative siblings \cite{zimmerer2019unsupervised,https://doi.org/10.48550/arxiv.1806.04972} for the \ac{uad} task, many recent CNN-based \ac{uad} networks were proposed~\cite{han2021madgan,tian2021constrained}. Regardless of their accomplishment, these networks have a weakness in learning long-range spatial dependencies, which can significantly impact segmentation performance for difficult tasks. 
\newline 
\textbf{Combining \ac{cnn}s with self-attention mechanisms:} Various studies have attempted to integrate self-attention mechanisms into \acp{cnn} by modeling global interactions of all pixels based on the feature maps. For instance, \cite{wang2018non} designed a non-local operator, which can be plugged into multiple intermediate convolution layers. Furthermore, \cite{schlemper2019attention} proposed additive attention gate modules built upon the \ac{ae} u-shaped architecture and integrated into the skip-connections. \cite{pinaya2021unsupervised} combined the latent discrete representation from the VQ-VAE model with an auto-regressive Transformer to obtain the likelihood of each latent variable value.
\newline 
\textbf{Vision Transformers:} 
For several computer vision tasks, \acp{vit} have rapidly acquired traction. By large-scale pretraining and fine-tuning of a basic Transformer architecture, \cite{https://doi.org/10.48550/arxiv.2010.11929} achieved \ac{sota} on ImageNet classification by directly applying Transformers with global self-attention to full-size images. 
In object detection, end-to-end Transformer-based models have shown prominence on several benchmarks \cite{carion2020end,dai2021dynamic}.
A few efforts \cite{hatamizadeh2022unetr,yan2022after} have explored the possibility of using Transformers for image segmentation tasks.
Furthermore, for medical image segmentation, \cite{valanarasu2021medical} proposed an axial fusion of Transformer with \ac{unet}. 
Recently, \cite{liu2021swin,wang2021pyramid} proposed hierarchical vision Transformers with varying resolutions and spatial embeddings.
For instance, instead of applying self-attention globally, \cite{parmar2018image} used it just in local neighborhoods for each query pixel. 
\cite{https://doi.org/10.48550/arxiv.1904.10509} proposed Sparse Transformers, which employ scalable approximations to global self-attention.

\section{Methodology}
We followed the recommended deep generative representation learning methodology, which is mentioned in \cite{baur2018deep} and illustrated in Fig.~\ref{fig1} (Appendix~\ref{appA}), to model the distribution of the healthy brain. This methodology should allow the model to fully reconstruct healthy brain anatomy while failing to reconstruct anomalous lesions in diseased brain MR images.
Anomalies are found and delimited by computing the pixel-wise $\mathcal{L}_{1}$-distance between an input image $\mathbf{x}$ and its reconstruction $\hat{\mathbf{x}}=h_{\phi}\left(g_{\theta}(\mathbf{x})\right)$, where $g_{\theta}$ and $h_{\phi}$ are, respectively, parametric functions of encoder-decoder networks of parameters $\theta$ and~$\phi$
\begin{equation}
\label{eq0}
\mathcal{L}_{R e c}^{\phi, \theta}(\mathbf{x}, \hat{\mathbf{x}}) = \mathcal{L}_{1}(\mathbf{x}, \hat{\mathbf{x}}).
\end{equation}
Then, we post-process the resulting residual image to get a binary segmentation.
This method uses healthy data to create a probability density estimate of the input data specified by the uniformity of the landscape.
Pathological traits are subsequently registered as deviations from normality, eliminating the need for labels or anomalous cases in training.

\subsection{Anomaly Segmentation}
First, we train our model to reconstruct healthy preprocessed input samples. This trained model is used to predict samples from anomalous preprocessed data. Then, we compute the residual or pixel-wise discrepancy between the input and the reconstruction. Fig.~\ref{fig_process} (Appendix~\ref{appA}) illustrates this process.
Assume that we are working on one channel (${\mathrm{C}}$=1), ${\mathbf{x} \mathrm{\in \mathcal{R}^{H \times W}}}$ is the input sample, ${\mathbf{\hat{x}} \mathrm{\in \mathcal{R}^{H \times W}}}$ is the reconstruction (prediction), ${\mathbf{r} \mathrm{\in \mathcal{R}^{H \times W}}}$ is the residual, ${\mathbf{l} \mathrm{ \in \{0,1\}^{H \times W} }}$ is the label relative to the ground truth segmentation of ${\mathbf{x}}$ and ${\mathbf{b_{mask}} \mathrm{ \in \{0,1\}^{H \times W}}}$ is the brain mask of the corresponding input ${\mathbf{x}}$.
We calculate the residual ${\mathbf{r}}$ which is a pixel-wise multiplication of ${\mathbf{b_{mask}}}$ with the difference between ${\mathbf{x}}$ and ${\mathbf{\hat{x}}}$.
\begin{equation}
\label{eq1}
\mathbf{r} = \mathbf{b_{mask}} \odot (\mathbf{x} - \hat{\mathbf{x}}).
\end{equation}

\subsection{Proposed Architectures}
\textbf{Basic Transformer \ac{ae}} is the first considered architecture, whose overview is depicted in Fig.~\ref{btae} (Appendix~\ref{appB}). This architecture is the simplest one intended to understand the behavior of Transformers alone and see the real effect of the self-attention mechanism and the added value of the global feature extractions on the reconstructions. Introduced in \cite{vaswani2017attention}, this \ac{btae} receives as input a 1D sequence of token embeddings.
To handle 2D images, we reshape the image ${\mathbf{x} \mathrm{\in \mathcal{R}^{H \times W \times C} }}$ into a sequence of flattened 2D patches ${\mathbf{x_{p}} \mathrm{\in \mathcal{R}^{N \times P^2 \times C}}}$, as proposed in the original \ac{vit} \cite{https://doi.org/10.48550/arxiv.2010.11929}, where (H, W) is the resolution of the original image, ${\mathrm{C}}$ is the number of channels, (P, P) is the resolution of each image patch, and ${\mathrm{N = \frac{H \times W}{P^2}}}$ is the resulting number of patches, which also serves as the effective input sequence length for the Transformer. 
The Transformer uses constant latent vector size ${\mathrm{K}}$ through all of its layers, so we flatten the patches and map to ${\mathrm{K}}$ dimensions with a linear projection. We refer to the output of this projection as the patch embeddings. 
Finally, we have another linear projection layer to project back embeddings into the initial dimension and a patch expanding layer to reconstruct the image with the same size $(H, W, C)$ as the input.
\begin{wrapfigure}{r}{0.43\textwidth}
  \vspace{-10mm}
  \begin{center}
    \includegraphics[width=\linewidth]{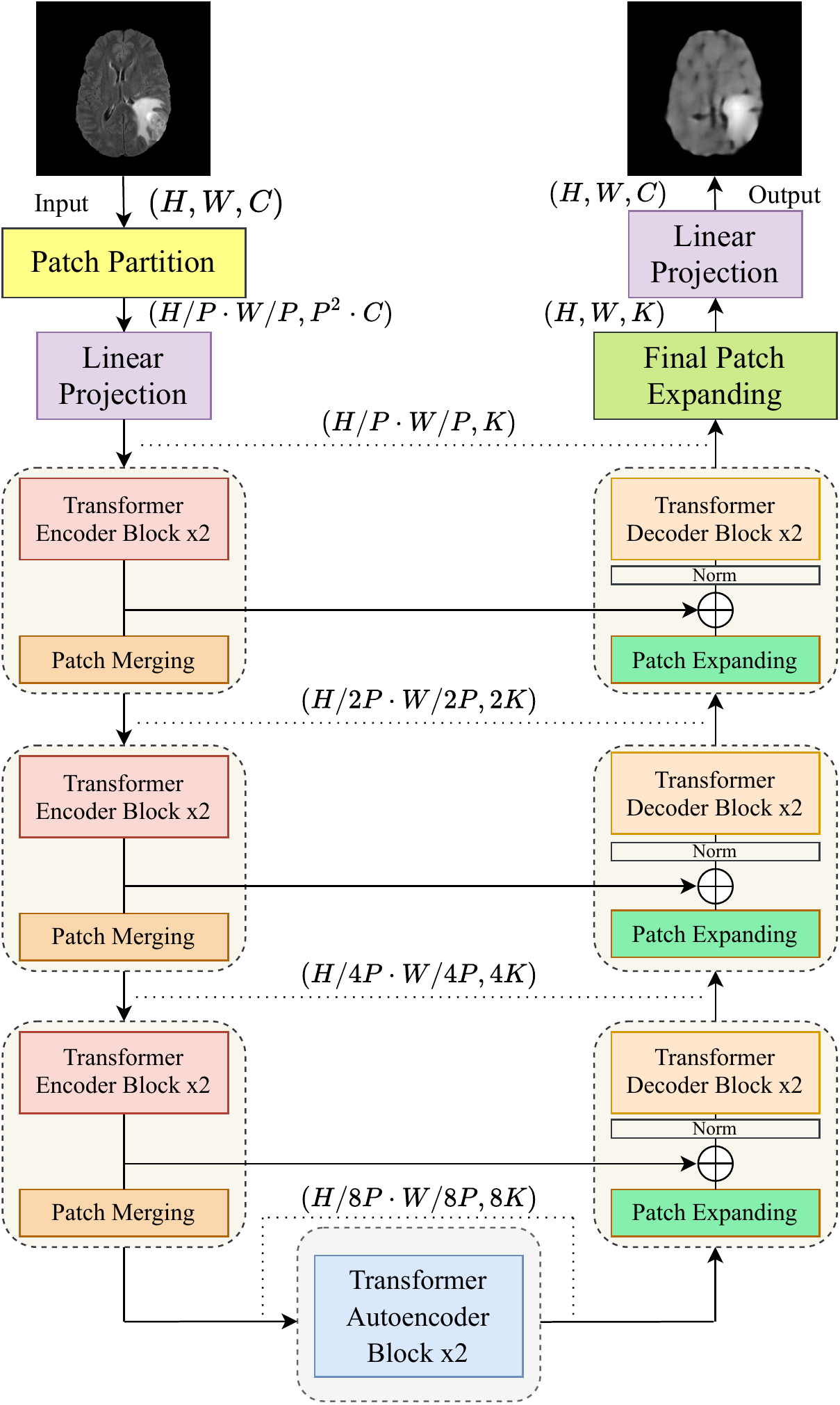}
  \end{center}
  \vspace{-6mm}
  \caption{\small Hierarchical Transformer Autoencoder with Skip connections (H\_TAE\_S).}
  \label{htaes}
  \vspace{-6mm}
\end{wrapfigure}
\textbf{Convolution inside Transformer \ac{ae}}: Based on the results of \ac{btae}, we found that the potential anomalous regions are recovered on the reconstructed images with the same intensity levels even after training on the healthy dataset due to the type of features extracted by Transformers (only global features).
From these results and inspired by the work in \cite{10.1007/978-3-030-87193-2_2}, we decided to use the complementary of Transformers and \acp{cnn} to improve the reconstruction capability of our \ac{btae}, and then, going deeper inside the Transformer bottleneck to combine local and global features by adding a Convolutional \ac{ae}.
Different from the methods mentioned in \cite{wang2021pyramid,zheng2021rethinking}, where they modeled hybrid \ac{cnn}-Transformer architectures, we propose a Transformer \ac{ae} network with \ac{cnn} \ac{ae} inside the bottleneck for the anomaly segmentation task. 
Two types of Convolutional \ac{ae}s were considered (see Appendix~\ref{appB}): \ac{sctae} with a spatial bottleneck (Fig.~\ref{sctae}) and \ac{dctae} with a dense bottleneck (Fig.~\ref{dctae}). These two convolutional autoencoders were implemented using a unified \ac{cnn} architecture. This unified architecture was also used for the benchmark models and thus having an accurate comparison as previously done in the comparative study in \cite{baur2021autoencoders}.

\textbf{Hierarchical Transformer \ac{ae}}: Inspired by Swin-Unet \cite{https://doi.org/10.48550/arxiv.2105.05537}, we presented a Hierarchical Transformer as \ac{ae} (Fig.~\ref{htaes}) and (Fig.~\ref{htae} in Appendix~\ref{appB}), with and without skip connections respectively, for gradually learning hierarchical features while reducing the computation complexity of standard Transformers for the medical image segmentation task.
Down/Up-sampling will replace the local features retrieved by \acp{cnn}, and when combined with the self-attention provided by Transformer layers, we get a fully Transformer architecture that is highly promising for the medical anomaly segmentation challenges.
For additional information see the Appendices~\ref{applayers} and \ref{appC}.

\section{Experiments}
\subsection{Implementation Details}
We implemented all models in TensorFlow, and the experimental study was carried out on a GPU cluster including two RTX 8000 devices (see Appendix~\ref{appG} for complexity analysis). All models were trained on the same OASIS-3~\cite{lamontagne2019oasis} training-set for 50 epochs with batch-size 12, using the ADAM optimizer, with parameters $\beta_1=0.9$, $\beta_2=0.999$ and $lr={10^{-5}}$, and \ac{mae} as loss function. We use the best validation score as a stopping criterion to save the best model. The output of all models is subject to the same post-processing: assuming that only positive pixels were kept and the others were set to zero, we post-process the residual ${\mathbf{r}}$ (Eq.~\eqref{eq1}) to obtain the binary mask ${\mathbf{m}}$ (Eq.~\eqref{eq2}) by applying a 2D median filter $\mathbf{F}$ with kernel $k=5$, and then applying squashing operator $\mathbf{S}$ to squash the 1\% lower intensities. Finally, we evaluate our segmentation according to the label $\mathbf{l}$.
\begin{equation}
\label{eq2}
\mathbf{m} = \mathbf{S}_{1\%} \left ( \mathbf{F}_{k=5} \left (\max(\mathbf{r}, 0) \right )\right )
\end{equation}

\begin{wraptable}{r}{7cm}
    \centering
    \vspace{-12mm}
    \caption{Considered datasets.} \label{tab_1}
    \begin{tabular}{@{}lcc@{}}
    \toprule
    Dataset  & Patients/Cohort & Train/Val/Test \\ 
    \midrule
    OASIS-3 & 574/Healthy & 16896/1280/4805 \\
    BraTS-20 & 31/GB & -/-/4805  \\
    MSLUB & 18/MS & -/-/3078 \\
    \bottomrule
    \end{tabular}%
    \vspace{-7mm}
\end{wraptable}
\subsection{Datasets}
We based our study on the human brain \ac{mr} scans with FLAIR modality (see Appendix~\ref{flairjustif} for the modality use justification), different planes, NIfTI format, and (H, W, C)=(256, 256, 1) sample size. More details can be found in Table~\ref{tab_1}. We apply the same preprocessing to all datasets: - the skull has been stripped with ROBEX \cite{iglesias2011robust}, - the resulting images have been normalized into the range [0,1] and resized using an interlinear interpolation.

\subsection{Evaluation Metrics}
We use \ac{dsc}, with mean and standard deviation, and \ac{auprc} to evaluate segmentation accuracy in our experiments. We report the \ac{auroc} as well. In addition, we use \ac{ssim} to evaluate the reconstruction fidelity on healthy data.

\section{Results}
\subsection{Overall Performance}
Table~\ref{tab_7} describes the overall performance of our Transformer-based models compared to the \ac{sota} models on each of the BraTS, MSLUB, and healthy unseen OASIS datasets.
DC\_TAE and H\_TAE\_S outperform the \ac{sota} models on BraTS dataset, and DC\_TAE has comparable performance with the \ac{sota} models on MSLUB. Moreover, the H\_TAE\_S and SC\_TAE have the best reconstruction fidelity performance.
Considering the models in \cite{pinaya2021unsupervised}, we outperform, on BraTS, their Models 1 and 2 but not their Model 3, and on MSLUB, we outperform only the Model 1. It's a relative comparison because we didn't use the same training set.
Finally, the qualitative results can be found in (Appendix~\ref{appF}) with visual examples of the different reviewed models on BraTS and MSLUB datasets in Fig.~\ref{fig_Qresb} and Fig.~\ref{fig_Qresm}, respectively. For more results about the ablation study and the parameter tuning process, see (Appendix~\ref{appD}) and (Appendix~\ref{appE}), respectively.

\subsection{Quantitative Evaluations}
Considering the \ac{dsc} metric on MSLUB and BraTS test-sets, DC\_TAE achieves an overall \ac{dsc} of $0.337$ and outperforms the first and second top-ranked \ac{sota} models by $1.7\%$ and $3.6\%$, respectively. H\_TAE\_S achieves an overall \ac{dsc} of $0.339$ and surpasses the first and second top-ranked \ac{sota} models by $1.9\%$ and $3.8\%$, respectively.
Considering the \ac{auprc} metric on MSLUB and BraTS test-sets, DC\_TAE achieves an overall \ac{auprc} of $0.313$ and achieves better results than the first and second top-ranked \ac{sota} models by $6.4\%$ and $7.4\%$, respectively. H\_TAE\_S achieves an overall \ac{auprc} of $0.264$ and outperforms the first and second top-ranked \ac{sota} models by $1.5\%$ and $2.5\%$, respectively.
Considering the \ac{ssim} on healthy OASIS test-set, H\_TAE\_S achieves an \ac{ssim} score of $0.9958$ and surpasses the first and second top-ranked \ac{sota} models by $5.43\%$ and $11.97\%$, respectively. SC\_TAE achieves an \ac{ssim} score of $0.9784$ and exceeds the first and second top-ranked \ac{sota} models by $3.69\%$ and $10.23\%$, respectively.

\begin{table}[!htpb]
\centering  
\caption{Overall performance on the BraTS and MSLUB and the healthy unseen OASIS dataset. Top-1 and top-2 scores are highlighted in bold/underline and bold, respectively. For each model, we provide an estimate of its theoretically best possible DICE score ([\emph{DSC}]) on each dataset. Models 1, 2 and 3 proposed in \cite{pinaya2021unsupervised} correspond to VQ-VAE + Transformer, VQ-VAE + Transformer + Masked Residuals and VQ-VAE + Transformer + Masked Residuals + Different Orderings models, respectively. $^*$ Models have not been trained on the same training set as the benchmark and our proposed models.}

\label{tab_7}
\resizebox{\textwidth}{!}{%
\begin{tabular}{@{}l|lll|lll|l@{}}
\toprule
\multirow{2}{*}{\textbf{Approach}}  & \multicolumn{3}{c |}{{\bf BraTS}} & \multicolumn{3}{c |}{{\bf MSLUB}} & \multicolumn{1}{c}{{\bf OASIS}}  \\ 
& \textbf{AUROC $\uparrow$} & \textbf{AUPRC $\uparrow$} & \textbf{[\emph{DSC}] ($\mu \pm \sigma$) $\uparrow$} & \textbf{AUROC $\uparrow$} & \textbf{AUPRC $\uparrow$} & \textbf{[\emph{DSC}] ($\mu \pm \sigma$) $\uparrow$} & \textbf{SSIM $\uparrow$}\\ \midrule
AE (dense) \cite{baur2021autoencoders}   &   0.7648   &  0.3493   &  0.4584 $\pm$ 0.4017 & 0.8765   &  0.1299   &  0.1687 $\pm$ 0.2955 & 0.8758 \\
AE (spatial) \cite{baur2021autoencoders}  &   0.6211   &  0.0857   &  0.3646 $\pm$ 0.4031 & 0.7430   &  0.0347   &  0.1539 $\pm$ 0.2893 & 0.9415 \\
VAE \cite{baur2021autoencoders}       &   0.7733   &  0.3923   &  0.4663 $\pm$ 0.4041 &  0.8263   &  0.0863   &   \underline{\textbf{0.1751}} $\pm$ \underline{\textbf{0.2985}} & 0.8761\\
VQ-VAE  \cite{van2017neural}     &   0.6992   &  0.2549   &  0.4306 $\pm$ 0.3889 &  \underline{\textbf{0.8881}}   &    \underline{\textbf{0.2441}}   &  0.1724 $\pm$ 0.2952 & 0.8724 \\
B\_TAE (Ours)     &   0.6142   &  0.1221   &  0.4252 $\pm$ 0.4475 & 0.7272   &  0.0177   &  0.1490 $\pm$ 0.2747 & 0.9610 \\
DC\_TAE (Ours)   &    \underline{\textbf{0.7802}}     &  \textbf{0.4250}   &  \textbf{0.5017} $\pm$ \textbf{0.4056} &  \textbf{0.8861}   &  \textbf{0.2025}   &  \textbf{0.1728} $\pm$ \textbf{0.2971} & 0.8269 \\
SC\_TAE (Ours)   &   0.6847   &  0.2561   &  0.4552 $\pm$ 0.4180 & 0.6669   &  0.0133   &  0.1306 $\pm$ 0.2682 & \textbf{0.9784} \\
H\_TAE (Ours)    &   0.5472   &  0.0716   &  0.3593 $\pm$ 0.4020 & 0.5788   &  0.0374   &  0.1592 $\pm$ 0.2865 & 0.8987 \\
H\_TAE\_S (Ours) &   \textbf{0.7763}   &   \underline{\textbf{0.5086}}   &   \underline{\textbf{0.5296}} $\pm$ \underline{\textbf{0.4417}} & 0.6906   &  0.0209   &  0.1489 $\pm$ 0.2857 & \underline{\textbf{0.9958}} \\ 
\midrule
Model 1$^*$ \cite{pinaya2021unsupervised} & $-$ & $-$ & 0.431 & $-$ & $-$ & 0.097 & $-$ \\
Model 2$^*$ \cite{pinaya2021unsupervised} & $-$ & $-$ & \textbf{0.476} & $-$ & $-$ & \textbf{0.234} & $-$ \\
Model 3$^*$ \cite{pinaya2021unsupervised} & $-$ & $-$ & \underline{\textbf{0.759}} & $-$ & $-$ & \underline{\textbf{0.378}} & $-$ \\
\bottomrule
\end{tabular}%
}
\end{table}

\subsection{Discussion}
CNN-based \acp{ae} are known for creating blurry reconstructions. This blurry reconstruction may cause residuals with high values in areas of the image with high frequencies, implying that the model cannot efficiently discriminate between healthy hyper-intensity areas and anomalies. To avoid these areas being mislabelled as anomalous, we conducted our research on three different approaches. We begin at first with the B\_TAE architecture and find that modeling only global context yields reconstructions with anomalies. Then, we explored another approach and found that DC\_TAE achieves superior performance due to its hybrid architectural design, which not only encodes strong global context by treating, along an attention mechanism, the image features as sequences but also well utilizes the low-level \ac{cnn} features. The evaluation results indicate that DC\_TAE is a stable and powerful network to extract both short and long-range dependencies of \ac{mri} scans. Finally, in our H\_TAE\_S model, we used the spatial information and attention mechanism in a u-shaped architectural design. The extracted context features are fused with multi-scale features from the encoder via skip connections to complement the loss of spatial information caused by down-sampling. By observing the results of anomaly detection scores for the anomalies of different sizes, fully Transformer architecture, i.e., without convolutions, yields less performance than \ac{sota} models for small anomalies. Hence, we can say that local features and small anomalies are correlated. According to BraTS dataset, we can say that better reconstruction fidelity corresponds to the highest segmentation results, but this is not true in MSLUB dataset. To properly study the correlation between \ac{ssim} and segmentation metrics, we should do some more experiments on different test sets.

\section*{Conclusion}
In this paper, we have proposed five Transformer-based image reconstruction models for anomaly segmentation. The models are based on three global architectures: basic, convolution inside, and hierarchical Transformer AE. Extensive experiments show that DC\_TAE and H\_TAE\_S achieve better or comparable performance than the \ac{sota} on two representative datasets, especially for small and complex anomalies. However, Transformers are known for modeling pixel dependencies due to their attention mechanism and hence should be able to substitute anomalous pixels with pseudo-healthy ones. However, none of our models succeeded in reconstructing this way. Yet, we can address this weakness as an area of improvement. Finally, we believe our findings will lead to more research into the properties of Transformers for anomaly detection and their application to advanced medical tasks.

%
%

\appendix

\newpage
\section{The core concept behind the reconstruction-based method}
\label{appA}
\begin{figure}[htbp]
    \centering
    \includegraphics[width=0.88\linewidth]{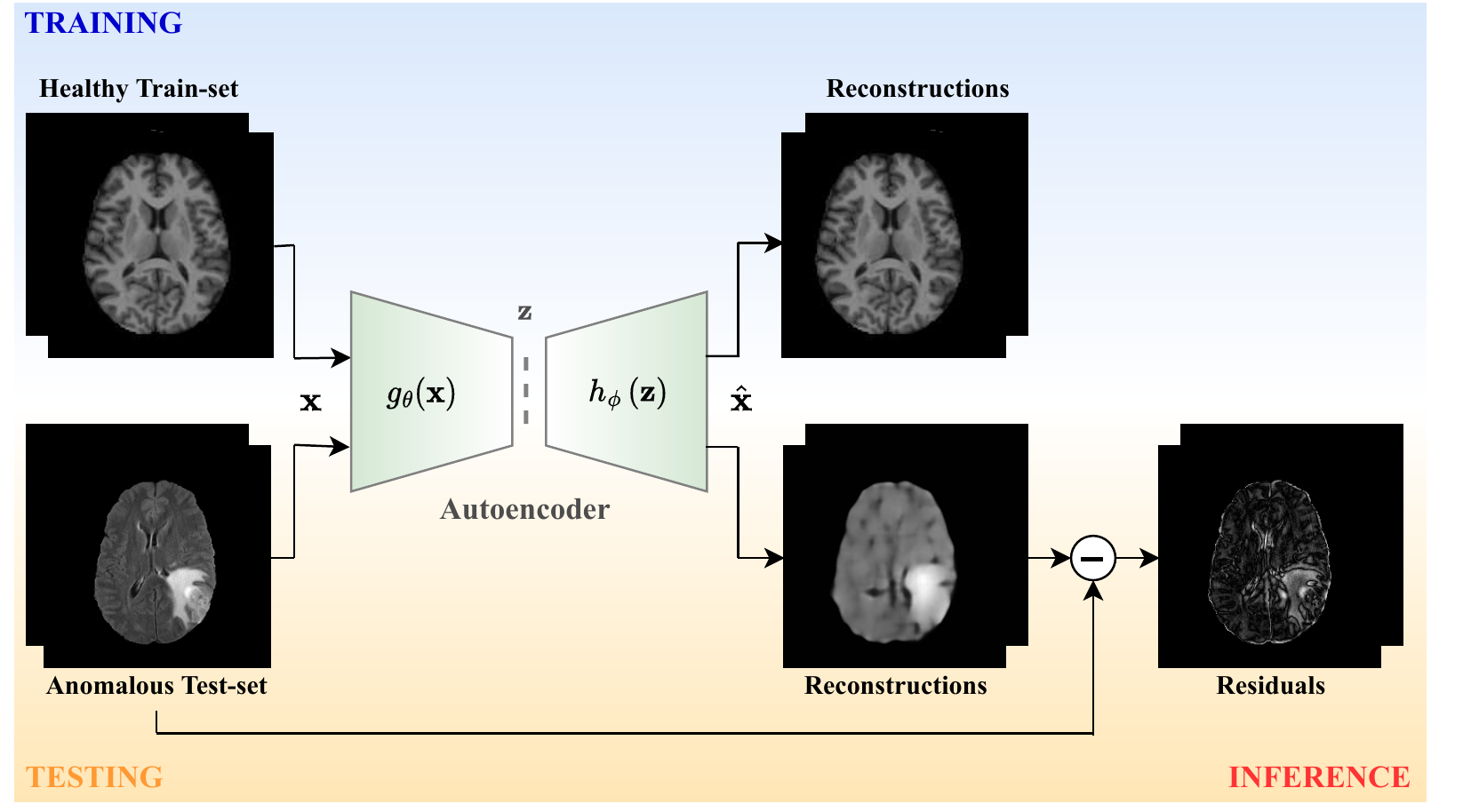}
    \caption{The concept of \ac{ae}-based anomaly segmentation: A) model training on healthy samples, B) model testing on anomalous samples, C) and segmentation of anomalies from reconstructions likely to carry an anomaly.}
    \label{fig1}
\end{figure}

\begin{figure}[htbp]
    \centering
    \includegraphics[width=0.86\linewidth]{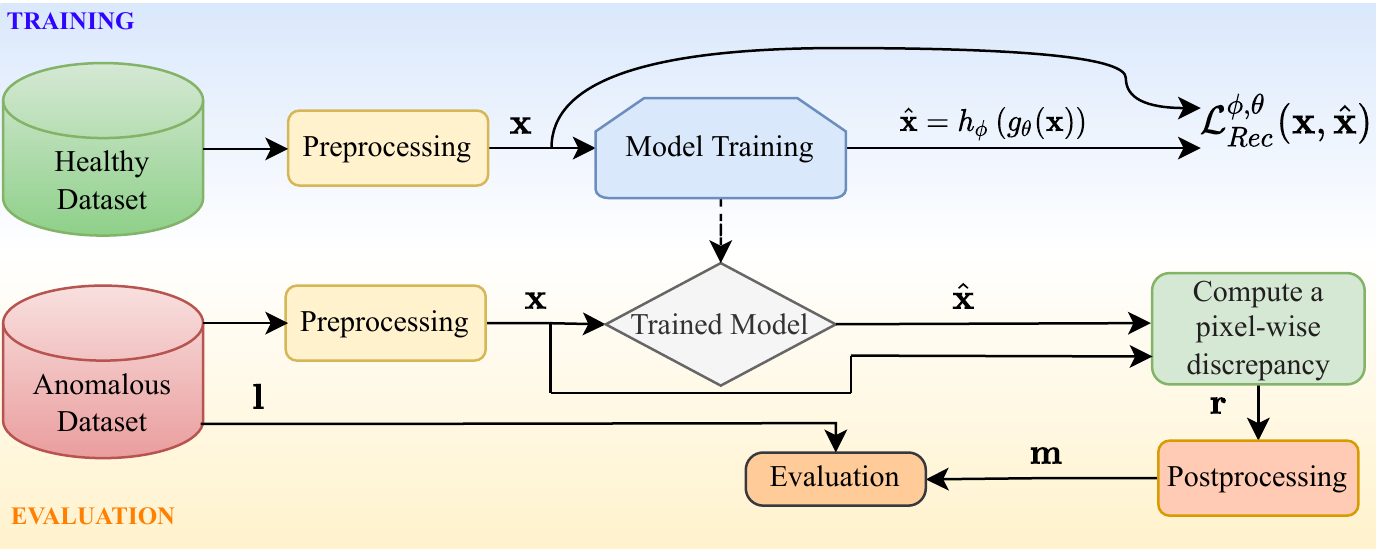}
    \caption{Unsupervised anomaly segmentation process, where $\mathbf{x}$:input, $\hat{\mathbf{x}}$:reconstruction, $\mathbf{r}$:residual, $\mathbf{l}$:label, $\mathbf{m}$:mask.}
    \label{fig_process}
\end{figure}

The core concept behind the reconstruction method is modeling healthy anatomy with unsupervised deep (generative) representation learning. Therefore, the method leverage a set of healthy MRI scans and learns to project and recover it from a lower dimensional distribution. See Fig.~\ref{fig1} and Fig.~\ref{fig_process} for more details.

\newpage
\section{Additional diagrams for the proposed Transformer based architectures}
\label{appB}

\subsection{Basic Transformer Autoencoder architecture}
\begin{figure}[htbp]
    \vspace{-8mm}
    \centering
    \includegraphics[width=0.3\linewidth]{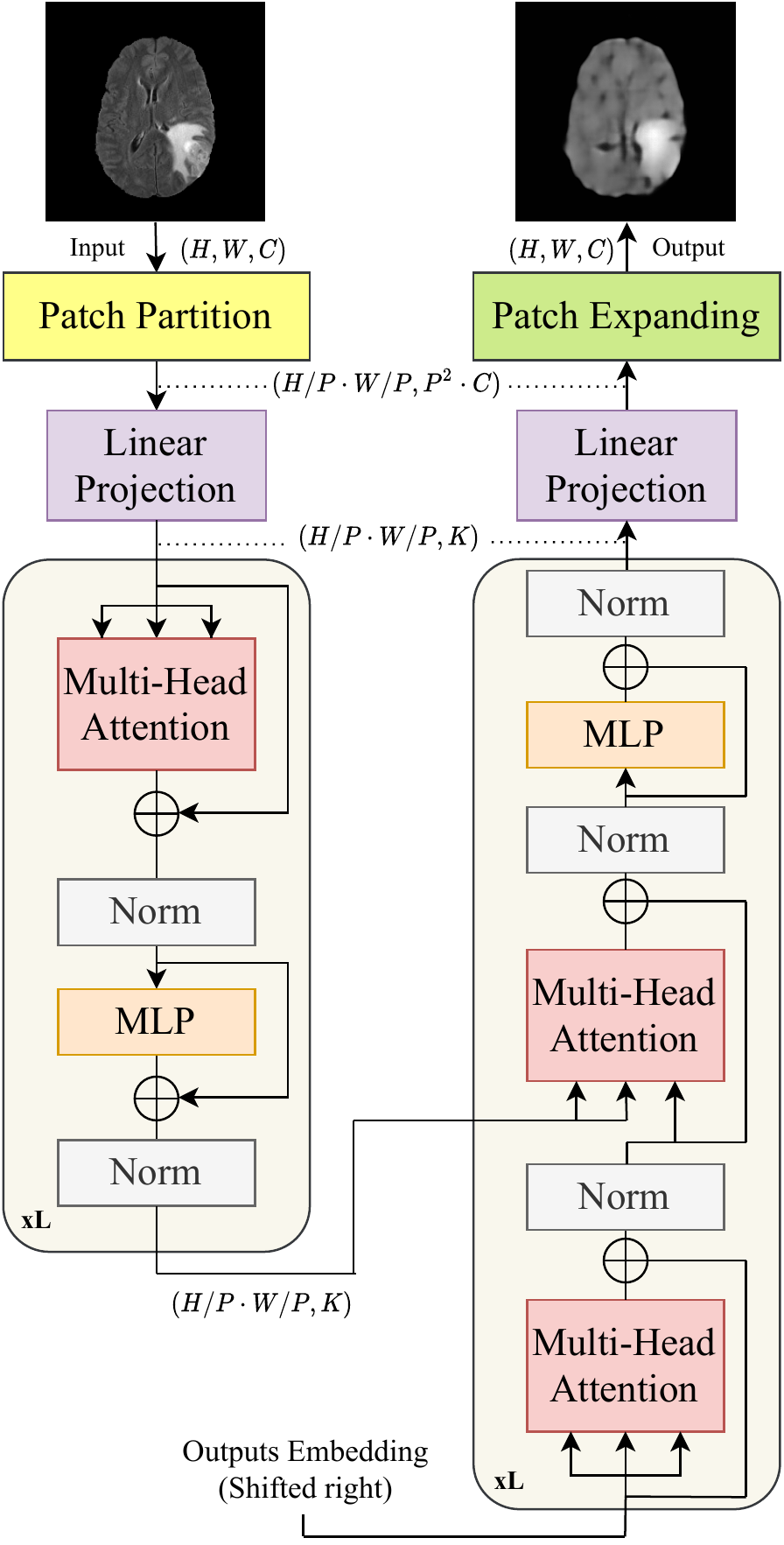}
    \caption{Basic Transformer Autoencoder (B\_TAE).}
    \label{btae}
\end{figure}

\vspace{-3mm}

\subsection{Spatial Convolutional Transformer Autoencoder architecture}
\begin{figure}[htbp]
    \vspace{-8mm}
    \centering
    \includegraphics[width=0.81\linewidth]{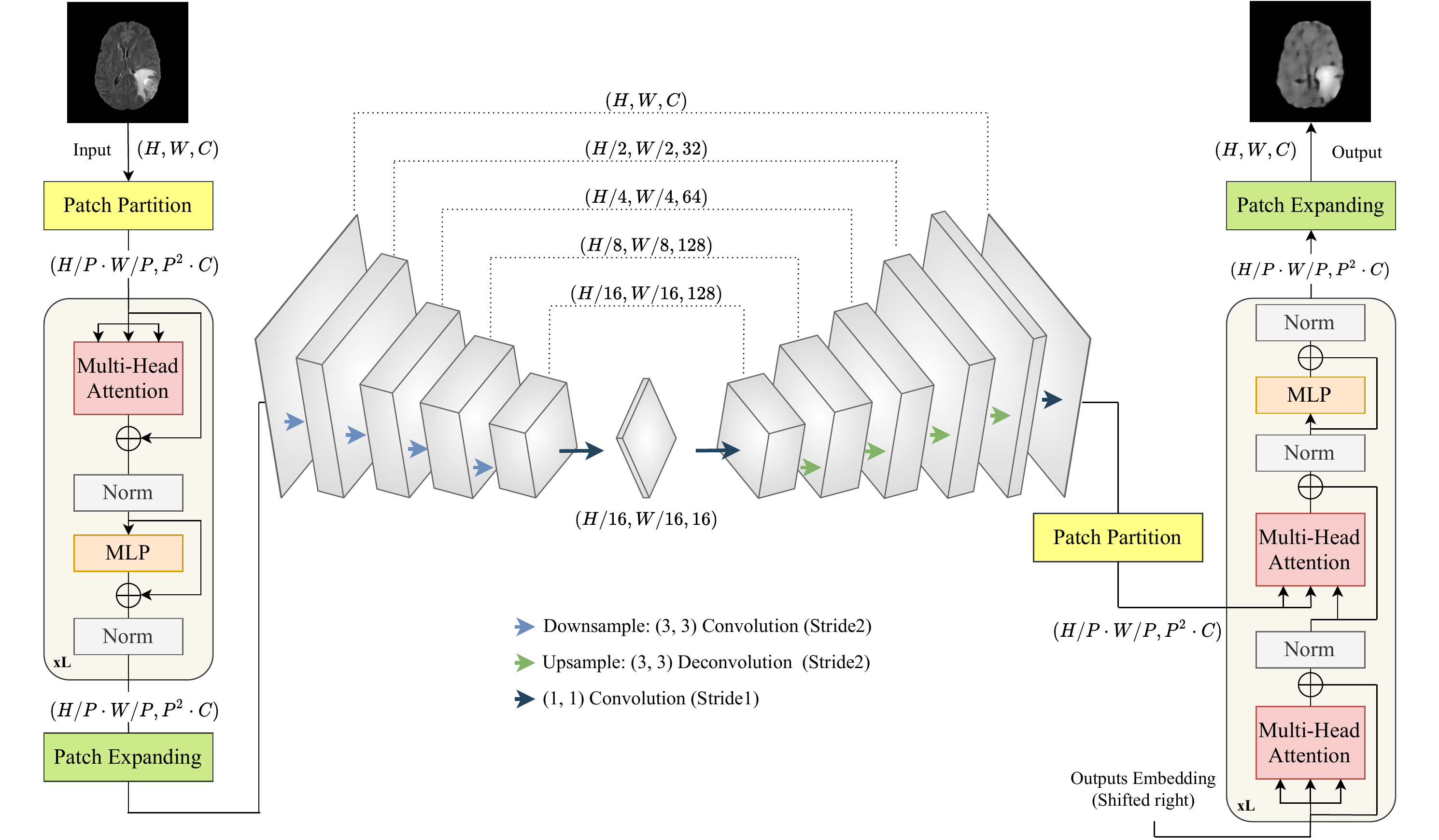}
    \caption{Spatial Convolutional Transformer Autoencoder (SC\_TAE).}
    \label{sctae}
\end{figure}

\newpage
\subsection{Dense Convolutional Transformer Autoencoder architecture}
\begin{figure}[htbp]
  \vspace{-8mm}
  \centering
  \includegraphics[width=0.81\linewidth]{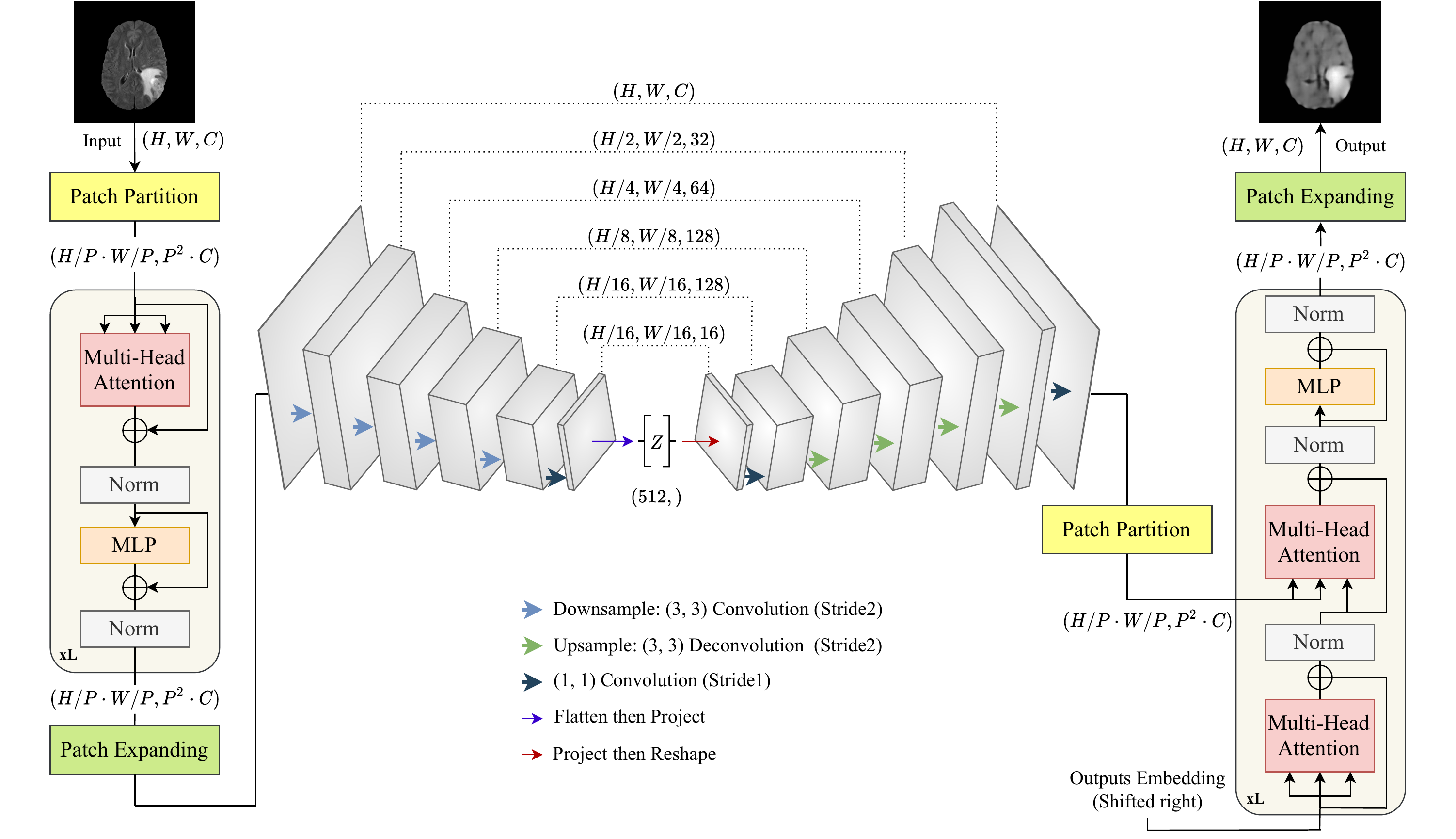}
  \caption{Dense Convolutional Transformer Autoencoder (DC\_TAE).}
  \label{dctae}
\end{figure}

\vspace{-3mm}

\subsection{Hierarchical Transformer Autoencoder architecture}
\begin{figure}[htbp]
  \vspace{-8mm}
  \centering
  \includegraphics[width=0.4\linewidth]{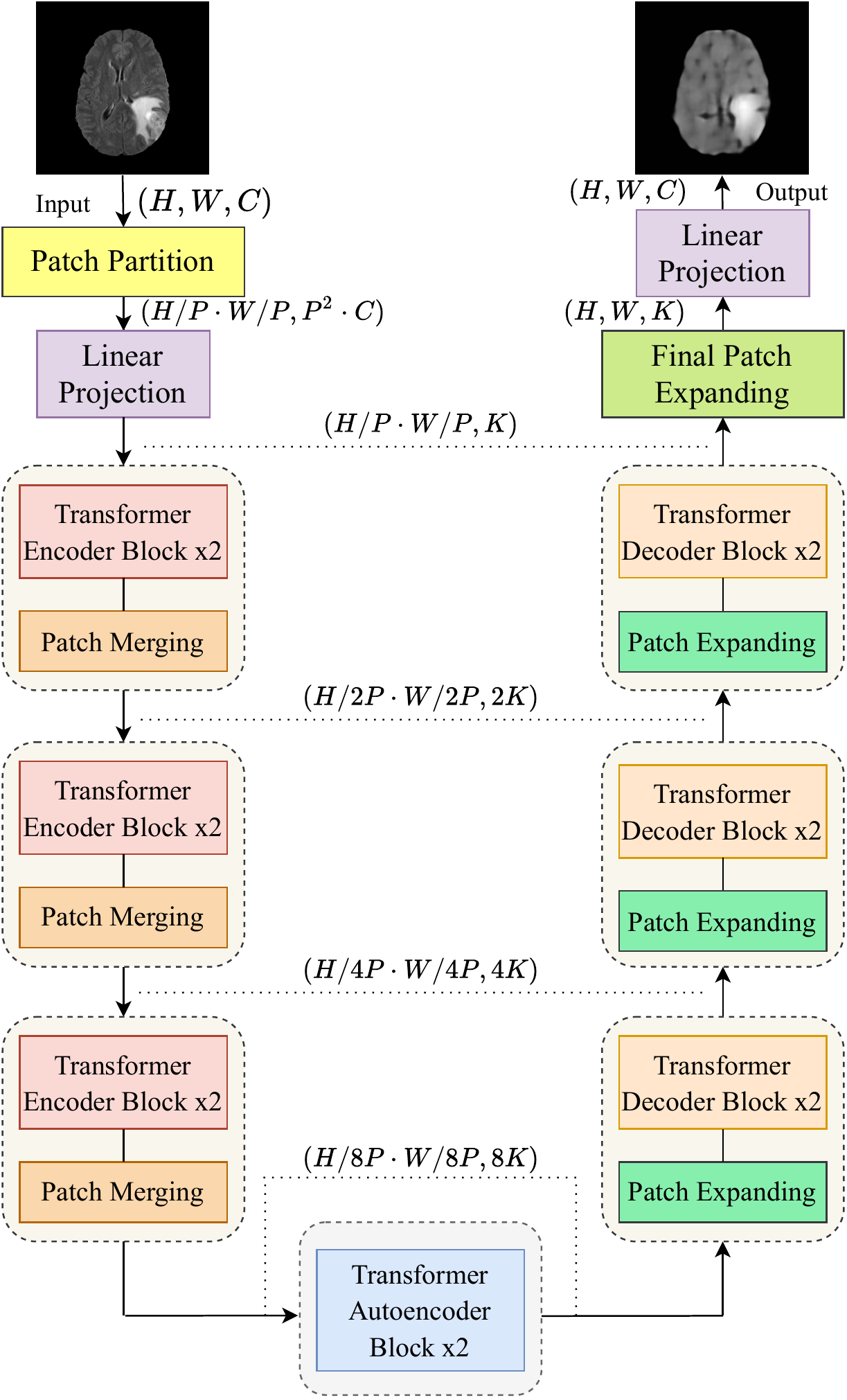}
  \caption{Hierarchical Transformer Autoencoder (H\_TAE).}
  \label{htae}
  \vspace{-11mm}
\end{figure}

\newpage
\section{Justification of working on FLAIR modality}
\label{flairjustif}
From a clinical perspective, FLAIR sequences are commonly used to identify and characterize imaging abnormalities according to their location, size, and extent in a wide range of pathologies and lesion manifestations \cite{duong2019convolutional}. FLAIR has been superior to other sequences, namely T2w, in detecting MS lesions, particularly those adjacent to the cerebral cortical gray matter \cite{bakshi2001fluid}. Therefore, we have constrained running our experiments on FLAIR sequences. This methodology would allow us to have a fair comparison with recent works, e.g., \cite{baur2018deep,zimmerer2019unsupervised,baur2021autoencoders,pinaya2021unsupervised}, among others.

\section{Additional information on the Hierarchical Transformer layers}
\label{applayers}

\begin{itemize}
\item \textbf{Patch Merging:} The patch merging layer decreases the number of tokens by a multiple of 2$\times$2 = 4 (2 down-sampling of resolution) and increases the feature dimension to half of the input dimension accordingly.
\item \textbf{Patch Expanding:} The patch expanding layer expands the number of tokens by a multiple of 2$\times$2 = 4 (2 up-sampling of resolution) and reduces the feature dimension to half of the input dimension accordingly.
\end{itemize}

\section{Contribution of the Transformer blocks in each of the proposed architectures}
\label{appblocks}
\begin{table}[!htb]
\centering
\caption{Role of the Transformer blocks according to each model.} \label{tab_cont}
\begin{tabular}{@{}l|ccccc@{}}
\toprule
\textbf{Role} & B\_TAE  & DC\_TAE & SC\_TAE & H\_TAE & H\_TAE\_S \\
\midrule
\begin{tabular}[c]{@{}l@{}}Encodes strong global context\\ by treating, along an attention\\ mechanism, the image features\\ as sequences\end{tabular} & \Checkmark    &  \Checkmark   &   \Checkmark   &   \XSolidBrush   &   \XSolidBrush\\
\midrule
\begin{tabular}[c]{@{}l@{}}Captures short/long-range spatial\\ correlations by combining the self-\\attention mechanism with down\\ and up-sampling\end{tabular} & \XSolidBrush   & \XSolidBrush   &   \XSolidBrush  &   \Checkmark  &   \Checkmark \\ 
\bottomrule
\end{tabular}%
\end{table}

%
\newpage
\section{Architecture specificities}
\label{appC}
We compared our five \ac{tae} models with \ac{sota} models as AE(dense), AE(spatial), \ac{vae} and \ac{vqvae} using the same training configurations and a unified \ac{cnn} architecture \cite{baur2021autoencoders} for all benchmark models. See Table~\ref{tab_2} for additional information about bottleneck size and intermediate dimension.
\begin{table}[!htpb]
\centering
\caption{Bottleneck sizes for each of the benchmark and proposed models.} \label{tab_2}
\resizebox{\textwidth}{!}{%
\begin{tabular}{@{}lcccccccc@{}}
\toprule
Model            & AE (dense) & AE (spatial) & VAE & VQ-VAE & B\_TAE & DC\_TAE & SC\_TAE & H\_TAE(\_S) \\ \midrule
Bottleneck       & (512,) & (16, 16, 16) & (512,) & (512,) & (256, 96) & (512,) & (16, 16, 16) & (1024, 768)\\ \bottomrule
\end{tabular}%
}
\end{table}

\section{Ablation study}
\label{appD}
Considering that \ac{btae} is a model deduced from the ablation of the AE(dense) and AE(spatial) respectively inside the \ac{dctae} and \ac{sctae} bottlenecks and that the H\_TAE is an ablation of the H\_TAE\_S's skip connections, we can interpret the following Tables~\ref{tab_3} and \ref{tab_4} as an ablation study results on the BraTS and MSLUB datasets, respectively.
Therefore, we can conclude from these results that the two models \ac{dctae} and H\_TAE\_S are globally the best-performing methods on average on the two considered test sets. 
\begin{table}[!htbp]
\centering
\caption{Ablation study results on the BraTS dataset.} \label{tab_3}
\begin{tabular}{@{}llll@{}}
\toprule
\textbf{Approach}   & \textbf{AUROC $\uparrow$} & \textbf{AUPRC $\uparrow$} & \textbf{[\emph{DSC}] ($\mu \pm \sigma$) $\uparrow$} \\ \midrule
B\_TAE     &   0.6142   &  0.1221   &  0.4252 $\pm$ 0.4475 \\
DC\_TAE    &   \textbf{0.7672}   &  \textbf{0.3498}   &  \textbf{0.4603} $\pm$ \textbf{0.4022} \\
SC\_TAE    &   0.6847   &  0.2561   &  0.4552 $\pm$ 0.4180 \\
H\_TAE     &   0.5472   &  0.0716   &  0.3593 $\pm$ 0.4020 \\
H\_TAE\_S  &   \underline{\textbf{0.7763}}   &  \underline{\textbf{0.5086}}   & \underline{\textbf{0.5296}} $\pm$ \underline{\textbf{0.4417}} \\ \bottomrule
\end{tabular}%
\end{table}

\vspace{-6mm}
\begin{table}[!htbp]
\centering
\caption{Ablation study results on the MSLUB dataset.} \label{tab_4}
\begin{tabular}{@{}llll@{}}
\toprule
\textbf{Approach}   & \textbf{AUROC $\uparrow$} & \textbf{AUPRC $\uparrow$} & \textbf{[\emph{DSC}] ($\mu \pm \sigma$) $\uparrow$} \\ \midrule
B\_TAE     &   \textbf{0.7272}   &  0.0177   &  0.1490 $\pm$ 0.2747 \\
DC\_TAE    &   \underline{\textbf{0.8745}}   &  \underline{\textbf{0.1609}}   &  \underline{\textbf{0.1631}} $\pm$ \underline{\textbf{0.2926}} \\
SC\_TAE    &   0.6669   &  0.0133   &  0.1306 $\pm$ 0.2682 \\
H\_TAE     &   0.5788   &  \textbf{0.0374}   &  \textbf{0.1592} $\pm$ \textbf{0.2865} \\
H\_TAE\_S  &   0.6906   &  0.0209   &  0.1489 $\pm$ 0.2857 \\ \bottomrule
\end{tabular}%
\end{table}

\newpage
\section{Parameters tuning}
\label{appE}
We considered, for the parameters tuning process, just for the two DC\_TAE and H\_TAE\_S models according to their performance on the last Tables~\ref{tab_3} and \ref{tab_4}. You can find the results of the parameters tuning process presented in Table~\ref{tab_5} for BraTS test set and Table~\ref{tab_6} for MSLUB test set. For the H\_TAE\_S we were able to test only two configurations because of the computational cost. Considering the results down below, we can conclude that the best configuration for the H\_TAE\_S is the one with (8: number of layers, 4: patch size, 4: number of heads) and for DC\_TAE is the one with (12: number of layers, 16: patch size, 8: number of heads).

\begin{table}[!htpb]
\centering
\caption{Parameters tuning results on the BraTS dataset.} \label{tab_5}
\resizebox{\textwidth}{!}{%
\begin{tabular}{@{}lllllll@{}}
\toprule
\textbf{Approach} & \textbf{Layers} & \textbf{Patches} & \textbf{Heads} & \textbf{AUROC $\uparrow$} & \textbf{AUPRC $\uparrow$} & \textbf{[\emph{DSC}] ($\mu \pm \sigma$) $\uparrow$} \\ \midrule
DC\_TAE    & 8  & 8  & 4    &   0.7755   &  0.3824   &  0.4821 $\pm$ 0.4058 \\
DC\_TAE    & 8  & 8  & 8    &   \underline{\textbf{0.8016}}   &  \textbf{0.5193}   &  0.4935 $\pm$ 0.4128 \\
DC\_TAE    & 8  & 16 & 4    &   0.7672   &  0.3498   &  0.4603 $\pm$ 0.4022 \\
DC\_TAE    & 8  & 16 & 8    &   0.7890   &  0.4673   &  \underline{\textbf{0.5060}} $\pm$ \underline{\textbf{0.4064}} \\
DC\_TAE    & 12 & 8  & 4    &   0.7773   &  0.3905   &  0.4714 $\pm$ 0.4052 \\
DC\_TAE    & 12 & 8  & 8    &   \textbf{0.8012}   &  \underline{\textbf{0.5411}}   &  0.4977 $\pm$ 0.4117 \\
DC\_TAE    & 12 & 16 & 8    &   0.7802   &  0.4250   & \textbf{0.5017} $\pm$ \textbf{0.4056} \\
\bottomrule
H\_TAE\_S  & 8  & 4  & 4    &   \textbf{0.7763}   &  \textbf{0.5086}  &  \textbf{0.5296} $\pm$ \textbf{0.4417} \\
H\_TAE\_S  & 8  & 4  & 8    &   0.5863   &  0.1346   &  0.3444 $\pm$ 0.3814 \\ \bottomrule
\end{tabular}%
}
\end{table}

\begin{table}[!htpb]
\centering
\caption{Parameters tuning results on the MSLUB dataset.} \label{tab_6}
\resizebox{\textwidth}{!}{%
\begin{tabular}{@{}lllllll@{}}
\toprule
\textbf{Approach} & \textbf{Layers} & \textbf{Patches} & \textbf{Heads} & \textbf{AUROC $\uparrow$} & \textbf{AUPRC $\uparrow$} & \textbf{[\emph{DSC}] ($\mu \pm \sigma$) $\uparrow$} \\ \midrule
DC\_TAE    & 8  & 8  & 4    &   \underline{\textbf{0.8909}}   &  0.0975   &  0.1562 $\pm$ 0.2931 \\
DC\_TAE    & 8  & 8  & 8    &   \textbf{0.8892}   &  0.0444   &  0.1556 $\pm$ 0.2920 \\
DC\_TAE    & 8  & 16 & 4    &   0.8745   &  0.1609   &  0.1631 $\pm$ 0.2926 \\
DC\_TAE    & 8  & 16 & 8    &   0.8765   &  0.1299   &  0.1687 $\pm$ 0.2955 \\
DC\_TAE    & 12 & 8  & 4    &   0.8776   &  \textbf{0.1812}   &  \underline{\textbf{0.1787}} $\pm$ \underline{\textbf{0.2958}} \\
DC\_TAE    & 12 & 8  & 8    &   0.8657   &  0.0329   &  0.1599 $\pm$ 0.2915 \\
DC\_TAE    & 12 & 16 & 8    &   0.8861   &  \underline{\textbf{0.2025}}   &  \textbf{0.1728} $\pm$ \textbf{0.2971} \\
\bottomrule
H\_TAE\_S  & 8  & 4  & 4    &   \textbf{0.6906}   &  \textbf{0.0209}   &  \textbf{0.1489} $\pm$ \textbf{0.2857} \\
H\_TAE\_S  & 8  & 4  & 8    &   0.5335   &  0.0046   &  0.1114 $\pm$ 0.2184 \\
\bottomrule
\end{tabular}%
}
\end{table}

\newpage
\section{Qualitative results}
\label{appF}
\begin{figure}[htbp]
  \centering
  \includegraphics[width=0.8\linewidth]{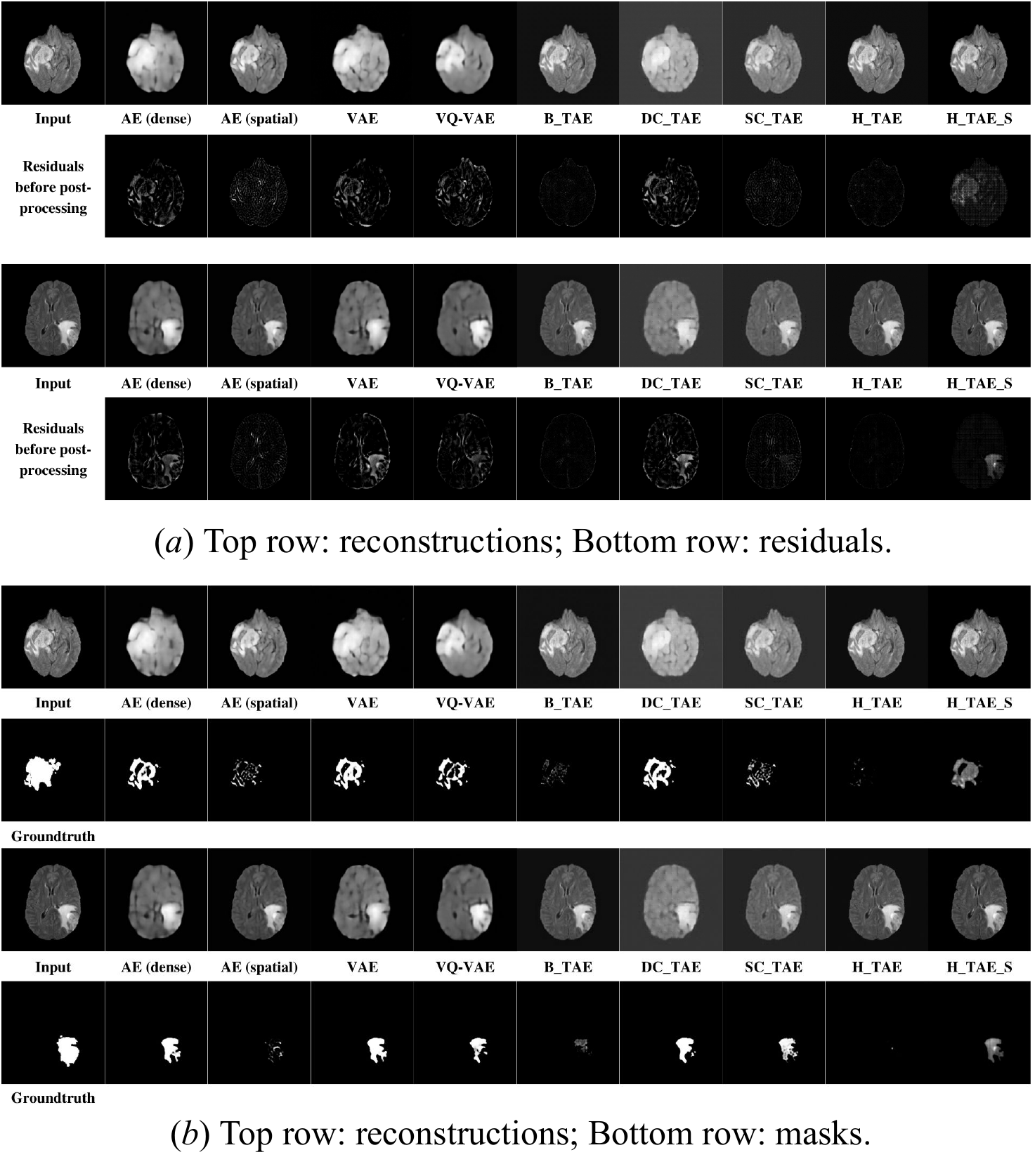}
  \caption{Visual examples of the different reviewed methods on BraTS dataset.}
  \label{fig_Qresb}
\end{figure}

\begin{figure}[htbp]
  \centering
  \includegraphics[width=0.8\linewidth]{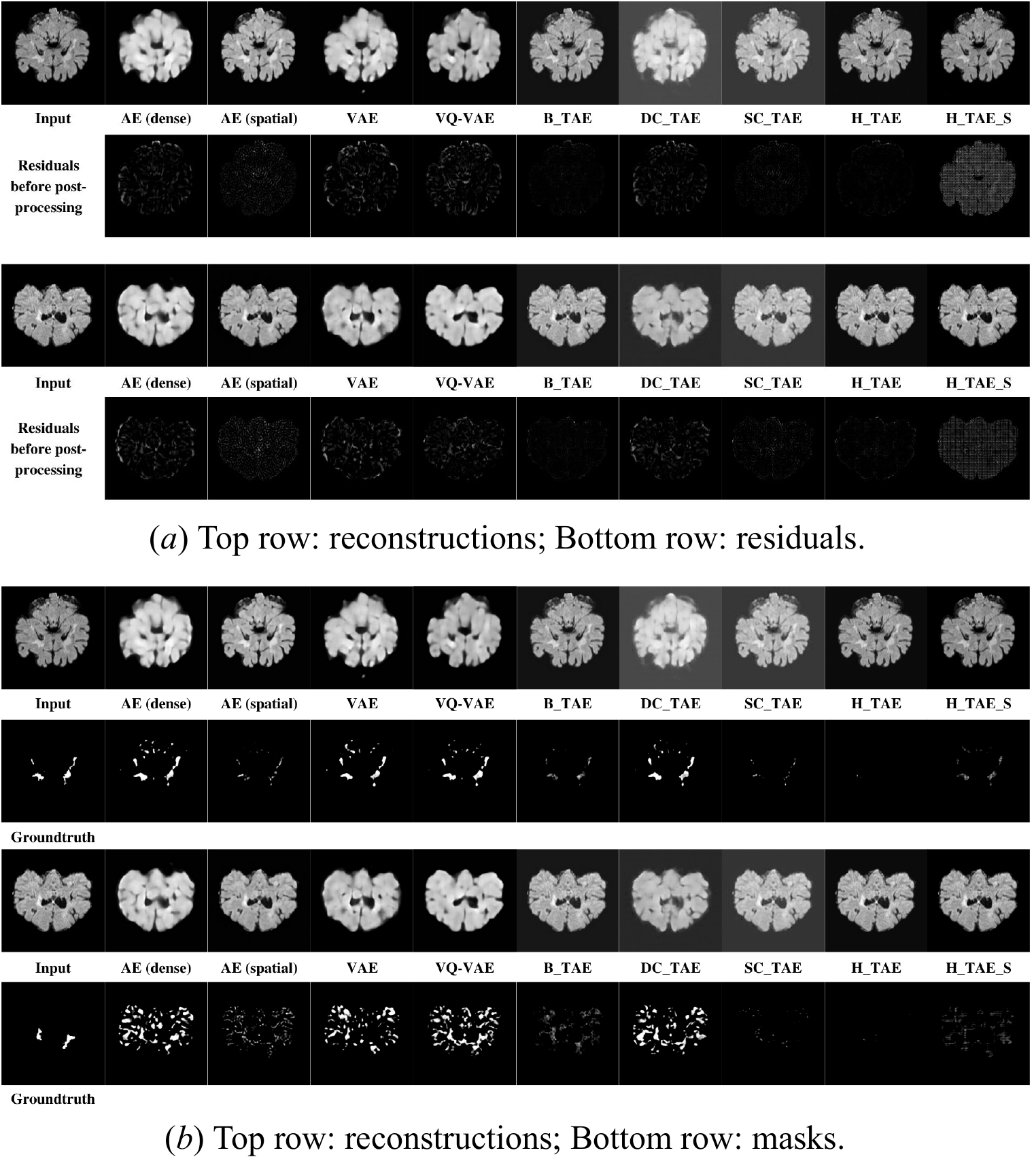}
  \caption{Visual examples of the different reviewed methods on MSLUB dataset.}
  \label{fig_Qresm}
\end{figure}

Qualitative brain segmentation comparisons are presented
in Fig.~\ref{fig_Qresb} and Fig.~\ref{fig_Qresm}. VAE, DC\_TAE, and H\_TAE\_S show improved segmentation performance of the \ac{gb} on BraTS dataset.
Segmenting the lesions in MSLUB dataset seems a little more difficult than detecting the \ac{gb} on BraTS, given the small size of the lesions.
Compared to the basic and spatial Transformer-based models, DC\_TAE exhibits higher boundary segmentation accuracy and performs better in capturing the fine-grained details of tumors.

\newpage
\section{Complexity analysis}
\label{appG}
\begin{figure}[htbp]
    \subfigure[BraTS  dataset]{%
    \label{fig_complexity_BraTS}
    \includegraphics[width= 0.5\linewidth]{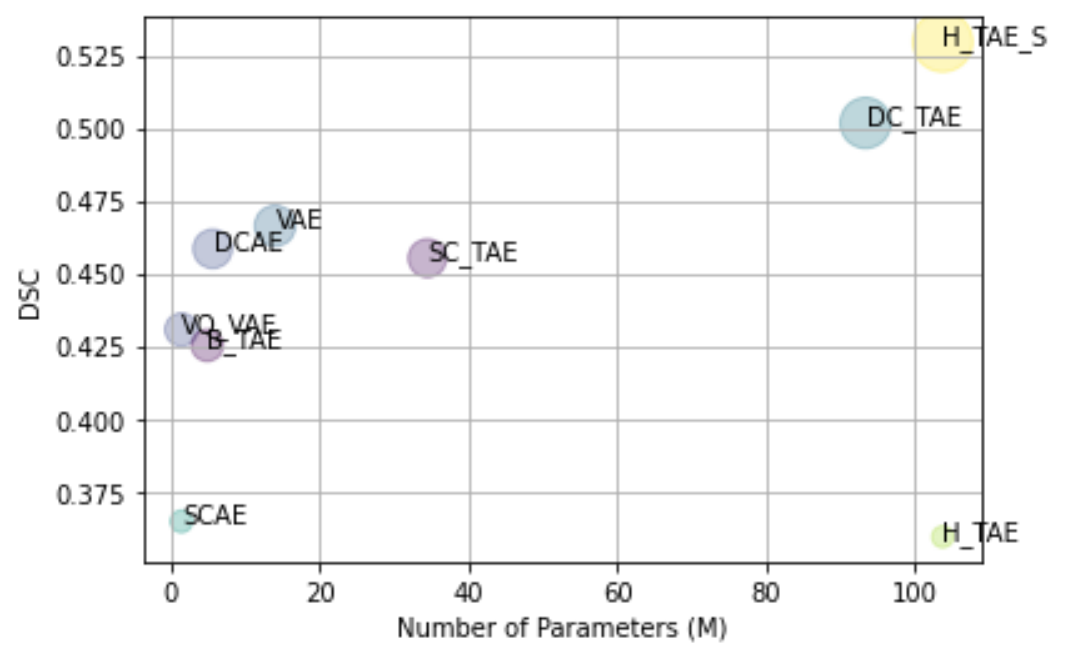}}
    \subfigure[MSLUB dataset]{%
    \label{fig_complexity_MSLUB}
    \includegraphics[width= 0.5\linewidth]{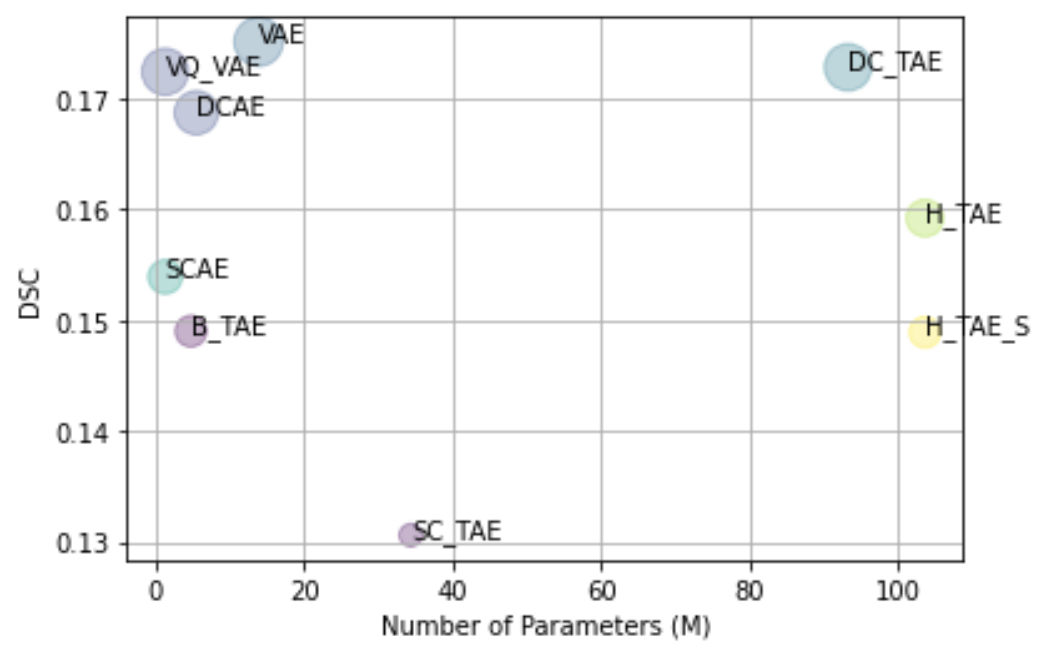}}
    \caption{Scatter graphs of the number of parameters $M$ in million vs Dice Similarity Coefficient (DSC) tested on BraTS and MSLUB datasets.}
    \label{fig_complexity}
\end{figure}

Scatter graphs of the number of parameters (in millions) versus the theoretically best possible DICE-score ([\emph{DSC}]) for our Transformer-based models and \ac{sota} models according to BraTS (left graph) and MSLUB (right graph) test-sets are displayed on Fig.~\ref{fig_complexity}.
We are aware of the model complexity that comes with the Transformers. However, we relax this complexity in our investigation, hoping to achieve a significant improvement over the \ac{sota} anomaly detection models. In the future, we would like to incorporate special algorithms into our models to improve performance and reduce computation complexity, as inspired by recent work \cite{liu2021swin} on innovative Transformer models for the image classification task.  

\begin{thebibliography}{8}
%
\bibitem{bakshi2001fluid}
Bakshi, R., Ariyaratana, S., Benedict, R. H., Jacobs, L. (2001). Fluid-attenuated inversion recovery magnetic resonance imaging detects cortical and juxtacortical multiple sclerosis lesions. Archives of neurology, 58(5), 742-748.
%
\bibitem{duong2019convolutional}
Duong, M. T., Rudie, J. D., Wang, J., Xie, L., Mohan, S., Gee, J. C., Rauschecker, A. M. (2019). Convolutional neural network for automated FLAIR lesion segmentation on clinical brain MR imaging. American Journal of Neuroradiology, 40(8), 1282-1290.
%
\bibitem{iglesias2011robust}
Iglesias, J. E., Liu, C. Y., Thompson, P. M., Tu, Z. (2011). Robust brain extraction across datasets and comparison with publicly available methods. IEEE transactions on medical imaging, 30(9), 1617-1634.
%
\bibitem{pinaya2021unsupervised}
Pinaya, W. H. L., Tudosiu, P. D., Gray, R., Rees, G., Nachev, P., Ourselin, S., Cardoso, M. J. (2021). Unsupervised brain anomaly detection and segmentation with transformers. arXiv preprint arXiv:2102.11650.
%
\bibitem{van2017neural}
Van Den Oord, A., Vinyals, O. (2017). Neural discrete representation learning. Advances in neural information processing systems, 30.
%
\bibitem{baur2018deep}
Baur, C., Wiestler, B., Albarqouni, S., Navab, N. (2018). Deep autoencoding models for unsupervised anomaly segmentation in brain MR images. In International MICCAI brainlesion workshop (pp. 161-169). Springer, Cham.
%
\bibitem{baur2021autoencoders}
Baur, C., Denner, S., Wiestler, B., Navab, N., Albarqouni, S. (2021). Autoencoders for unsupervised anomaly segmentation in brain MR images: a comparative study. Medical Image Analysis, 69, 101952.
%
\bibitem{zimmerer2019unsupervised}
Zimmerer, D., Isensee, F., Petersen, J., Kohl, S., Maier-Hein, K. (2019). Unsupervised anomaly localization using variational auto-encoders. In International Conference on Medical Image Computing and Computer-Assisted Intervention (pp. 289-297). Springer, Cham.
%
\bibitem{baur2021modeling}
Baur, C., Wiestler, B., Muehlau, M., Zimmer, C., Navab, N., Albarqouni, S. (2021). Modeling healthy anatomy with artificial intelligence for unsupervised anomaly detection in brain mri. Radiology: Artificial Intelligence, 3(3).
%
\bibitem{valanarasu2021medical}
Valanarasu, J. M. J., Oza, P., Hacihaliloglu, I., Patel, V. M. (2021). Medical transformer: Gated axial-attention for medical image segmentation. In International Conference on Medical Image Computing and Computer-Assisted Intervention (pp. 36-46). Springer, Cham.
%
\bibitem{hatamizadeh2022unetr}
Hatamizadeh, A., Tang, Y., Nath, V., Yang, D., Myronenko, A., Landman, B., Xu, D. (2022). Unetr: Transformers for 3d medical image segmentation. In Proceedings of the IEEE/CVF Winter Conference on Applications of Computer Vision (pp. 574-584).
%
\bibitem{yan2022after}
Yan, X., Tang, H., Sun, S., Ma, H., Kong, D., Xie, X. (2022). After-unet: Axial fusion transformer unet for medical image segmentation. In Proceedings of the IEEE/CVF Winter Conference on Applications of Computer Vision (pp. 3971-3981).
%
\bibitem{liu2021swin}
Liu, Z., Lin, Y., Cao, Y., Hu, H., Wei, Y., Zhang, Z., Guo, B. (2021). Swin transformer: Hierarchical vision transformer using shifted windows. In Proceedings of the IEEE/CVF International Conference on Computer Vision (pp. 10012-10022).
%
\bibitem{wang2021pyramid}
Wang, W., Xie, E., Li, X., Fan, D. P., Song, K., Liang, D., Shao, L. (2021). Pyramid vision transformer: A versatile backbone for dense prediction without convolutions. In Proceedings of the IEEE/CVF International Conference on Computer Vision (pp. 568-578).
%
\bibitem{carion2020end}
Carion, N., Massa, F., Synnaeve, G., Usunier, N., Kirillov, A., Zagoruyko, S. (2020). End-to-end object detection with transformers. In European conference on computer vision (pp. 213-229). Springer, Cham.
%
\bibitem{dai2021dynamic}
Dai, X., Chen, Y., Yang, J., Zhang, P., Yuan, L., Zhang, L. (2021). Dynamic detr: End-to-end object detection with dynamic attention. In Proceedings of the IEEE/CVF International Conference on Computer Vision (pp. 2988-2997).
%
\bibitem{wang2018non}
Wang, X., Girshick, R., Gupta, A., He, K. (2018). Non-local neural networks. In Proceedings of the IEEE conference on computer vision and pattern recognition (pp. 7794-7803).
%
\bibitem{schlemper2019attention}
Schlemper, J., Oktay, O., Schaap, M., Heinrich, M., Kainz, B., Glocker, B., Rueckert, D. (2019). Attention gated networks: Learning to leverage salient regions in medical images. Medical image analysis, 53, 197-207.
%
\bibitem{vaswani2017attention}
Vaswani, A., Shazeer, N., Parmar, N., Uszkoreit, J., Jones, L., Gomez, A. N., Polosukhin, I. (2017). Attention is all you need. Advances in neural information processing systems, 30.
%
\bibitem{parmar2018image}
Parmar, N., Vaswani, A., Uszkoreit, J., Kaiser, L., Shazeer, N., Ku, A., Tran, D. (2018). Image transformer. In International conference on machine learning (pp. 4055-4064). PMLR.
%
\bibitem{https://doi.org/10.48550/arxiv.1904.10509}
Child, R., Gray, S., Radford, A., Sutskever, I. (2019). Generating long sequences with sparse transformers. arXiv preprint arXiv:1904.10509.
%
\bibitem{https://doi.org/10.48550/arxiv.2010.11929}
Dosovitskiy, A., Beyer, L., Kolesnikov, A., Weissenborn, D., Zhai, X., Unterthiner, T., Houlsby, N. (2020). An image is worth 16x16 words: Transformers for image recognition at scale. arXiv preprint arXiv:2010.11929.
%
\bibitem{isensee2021nnu}
Isensee, F., Jaeger, P. F., Kohl, S. A., Petersen, J., Maier-Hein, K. H. (2021). nnU-Net: a self-configuring method for deep learning-based biomedical image segmentation. Nature methods, 18(2), 203-211.
%
\bibitem{siddique2021u}
Siddique, N., Paheding, S., Elkin, C. P., Devabhaktuni, V. (2021). U-net and its variants for medical image segmentation: A review of theory and applications. Ieee Access, 9, 82031-82057.
%
\bibitem{https://doi.org/10.48550/arxiv.1806.04972}
Chen, X., Konukoglu, E. (2018). Unsupervised detection of lesions in brain MRI using constrained adversarial auto-encoders. arXiv preprint arXiv:1806.04972.
%
\bibitem{bruno2015understanding}
Bruno, M. A., Walker, E. A., Abujudeh, H. H. (2015). Understanding and confronting our mistakes: the epidemiology of error in radiology and strategies for error reduction. Radiographics, 35(6), 1668-1676.
%
\bibitem{taboada2009anomaly}
Taboada-Crispi, A., Sahli, H., Hernandez-Pacheco, D., Falcon-Ruiz, A. (2009). Anomaly detection in medical image analysis. In Handbook of research on advanced techniques in diagnostic imaging and biomedical applications (pp. 426-446). IGI Global.
%
\bibitem{zheng2021rethinking}
Zheng, S., Lu, J., Zhao, H., Zhu, X., Luo, Z., Wang, Y., Zhang, L. (2021). Rethinking semantic segmentation from a sequence-to-sequence perspective with transformers. In Proceedings of the IEEE/CVF conference on computer vision and pattern recognition (pp. 6881-6890).
%
\bibitem{https://doi.org/10.48550/arxiv.2105.05537}
Cao, H., Wang, Y., Chen, J., Jiang, D., Zhang, X., Tian, Q., Wang, M. (2021). Swin-unet: Unet-like pure transformer for medical image segmentation. arXiv preprint arXiv:2105.05537.
%
\bibitem{10.1007/978-3-030-87193-2_2}
Zhang, Y., Liu, H., Hu, Q. (2021, September). Transfuse: Fusing transformers and cnns for medical image segmentation. In International Conference on Medical Image Computing and Computer-Assisted Intervention (pp. 14-24). Springer, Cham.
%
\bibitem{lamontagne2019oasis}
LaMontagne, P. J., Benzinger, T. L., Morris, J. C., Keefe, S., Hornbeck, R., Xiong, C., Marcus, D. (2019). OASIS-3: longitudinal neuroimaging, clinical, and cognitive dataset for normal aging and Alzheimer disease. MedRxiv.
%
\bibitem{menze2014multimodal}
Menze, B. H., Jakab, A., Bauer, S., Kalpathy-Cramer, J., Farahani, K., Kirby, J., Van Leemput, K. (2014). The multimodal brain tumor image segmentation benchmark (BRATS). IEEE transactions on medical imaging, 34(10), 1993-2024.

\bibitem{lesjak2018novel}
Lesjak, Ž., Galimzianova, A., Koren, A., Lukin, M., Pernuš, F., Likar, B., Špiclin, Ž. (2018). A novel public MR image dataset of multiple sclerosis patients with lesion segmentations based on multi-rater consensus. Neuroinformatics, 16(1), 51-63.
%
\bibitem{zhang2020multiple}
Zhang, H., Oguz, I. (2020). Multiple sclerosis lesion segmentation-a survey of supervised cnn-based methods. In International MICCAI Brainlesion Workshop (pp. 11-29). Springer, Cham.
%
\bibitem{han2021madgan}
Han, C., Rundo, L., Murao, K., Noguchi, T., Shimahara, Y., Milacski, Z. Á., Satoh, S. I. (2021). MADGAN: Unsupervised medical anomaly detection GAN using multiple adjacent brain MRI slice reconstruction. BMC bioinformatics, 22(2), 1-20.
%
\bibitem{tian2021constrained}
Tian, Y., Pang, G., Liu, F., Chen, Y., Shin, S. H., Verjans, J. W., Carneiro, G. (2021). Constrained contrastive distribution learning for unsupervised anomaly detection and localisation in medical images. In International Conference on Medical Image Computing and Computer-Assisted Intervention (pp. 128-140). Springer, Cham.
%
\bibitem{van2016pixel}
Van Oord, A., Kalchbrenner, N., Kavukcuoglu, K. (2016). Pixel recurrent neural networks. In International conference on machine learning (pp. 1747-1756). PMLR.
%
\end{thebibliography}
\end{document}